\documentclass{elsart}

\usepackage{psfig}

\usepackage{natbib}
\def\ergsec{\hbox{erg s$^{-1}$}}
\def\ergcm{\hbox{erg cm$^{-2}$ s$^{-1}$}}

\begin{document}

\runauthor{W. Pietsch, K. Bischoff, Th. Boller}


\begin{frontmatter}

\title{New NLS1 galaxies detected among RASS galaxies}

\author[mpe]{W. Pietsch}, \author[uni]{K. Bischoff}, \author[mpe]{Th. Boller}
\address[mpe]{Max-Planck-Institut f\"ur Extraterrestrische Physik, Garching, Germany}
\address[uni]{Universit\"ats-Sternwarte, G\"ottingen, Germany}

\begin{abstract}
Six Narrow-Line Seyfert 1 galaxies have been discovered in optical follow-up
spectroscopy of nuclei of galaxies selected for their high X-ray flux
in the ROSAT All Sky Survey. We give X-ray and optical parameters of the sources 
and show optical spectra.  

\end{abstract}


\end{frontmatter}


\section{Introduction}

Cross correlations of the ROSAT All Sky Survey (RASS) source catalogue with
the Catalogue of Principal Galaxies (Paturel et al. 1989) 
yielded 903 X-ray sources with galaxy
counterparts (Zimmermann et al., in preparation).
In several of them the X-ray emission can be attributed to
known Active Galactic Nuclei (AGN) or emission from hot gas in clusters of
galaxies, while in other galaxies the detected X-ray luminosity is in the
range expected for normal spirals and ellipticals following the results of
the {\it Einstein} observatory (Fabbiano et al. 1992).
Of special interest were X-ray bright galaxies for
which no reason for their X-ray brightness was known. They were followed up
by X-ray imaging with the ROSAT HRI and/or optical spectroscopy using the
2.2 m telescopes at Calar Alto and La Silla. In this program we detected more
than 100 previously unknown relatively nearby AGN of different types
(Pietsch et al. 1998; Bischoff et al., in preparation).
Here we present details on six newly discovered narrow line Seyfert 1 galaxies.

\section{X-ray and optical properties of newly detected NLS1}
In Table 1 we summarize the soft X-ray and optical properties of the newly
detected NLS1 from the sample. We have used the FWHM of H$\alpha < 2200$ km s$^{-1}$
to select NLS1 candidates. As the FWHM of H$\alpha$ is usually larger than
that of H$\beta$, our selection criterium is in agreement with the original
NLS1 definition of Osterbrock \& Pogge (1985) and Goodrich (1989). 
With the exception of CGCG 374-029, all the objects show strong Fe{\,\sc ii} 
emission in their optical spectra (see Fig. 1), supporting the NLS1 nature.
In the table we give redshift, FWHM of H$\alpha$, and the distance as determined
from the optical spectra.
  
In X-rays we give RASS count rates and hardness ratios. Count rates
have been converted to fluxes assuming
power law spectra with a photon index 2.3 and correcting for galactic absorption.
Luminosities have been computed assuming H$_0$ = 75 km s$^{-1}$ and $q_0$=0.5.

The preliminary census of NLS1 of our analysis of X-ray selected
nearby AGN is consistent with the NLS1 content of other X-ray selected AGN
samples.   

   \begin{table}
   {\small{
      \caption{Parameters of new NLS1 galaxies.}
         \label{table} 
         \begin{flushleft}
         \begin{tabular}{lrrrrrrrr}
            \hline
            \noalign{\smallskip}
Galaxy & \multicolumn{4}{c}{RASS} & Redshift & FWHM & D & L$_{\rm X}$  \\
            \noalign{\smallskip}
       & rate & HR1 & HR2 & f$_{\rm X}$ & cz & H$\alpha$ & & \\
            \noalign{\smallskip}
       & ($^{\rm \ast}$)& & & ($^{\rm \ast\ast}$) & \multicolumn{2}{c}{(km s$^{-1}$)} & (Mpc) &($^{\rm \ast\ast}$)\\
            \noalign{\smallskip}
            \hline
            \noalign{\smallskip}
KUG 1618+410 & 0.51 & -0.54 & +0.04 &  14 &  $8\,532\pm29$  & 2\,159 & 114 &  22 \\
CGCG 374-029 & 0.05 & +1.00 & +0.76 & 4.8 &  $4\,005\pm82$  & 2\,048 &  53 & 1.6 \\
CGCG 493-004 & 0.07 & +1.00 & +0.51 & 8.5 & $13\,042\pm46$  & 1\,410 & 174 &  31 \\
NGC 7158     & 0.29 & +0.03 & -0.15 &  19 &  $8\,275\pm30$  & 2\,100 & 110 &  28 \\
II ZW 177    & 0.25 & +0.04 & -0.45 &  20 & $24\,440\pm110$ & 1\,176 & 326 & 250 \\
ESO 407-17   & 0.22 & -0.79 & -0.55 & 7.9 & $27\,370\pm30$  & 1\,480 & 365 & 130 \\
            \noalign{\smallskip}
            \hline
            \noalign{\smallskip}
         \end{tabular}
{$^{\rm \ast}$: count rate in cts s$^{-1}$\\
$^{\rm \ast\ast}$: X-ray flux and luminosity in units of 10$^{-13}$\,\ergcm\ and 10$^{41}$\,\ergsec, 
respectively, corrected for Galactic foreground absorption}
         \end{flushleft}
   }}
   \end{table}


\begin{figure}[htb]
{
\hbox{\begin{minipage}{6.0cm}
      \vspace{0.5cm}
      \centerline{KUG 1618+410}
      \psfig{figure=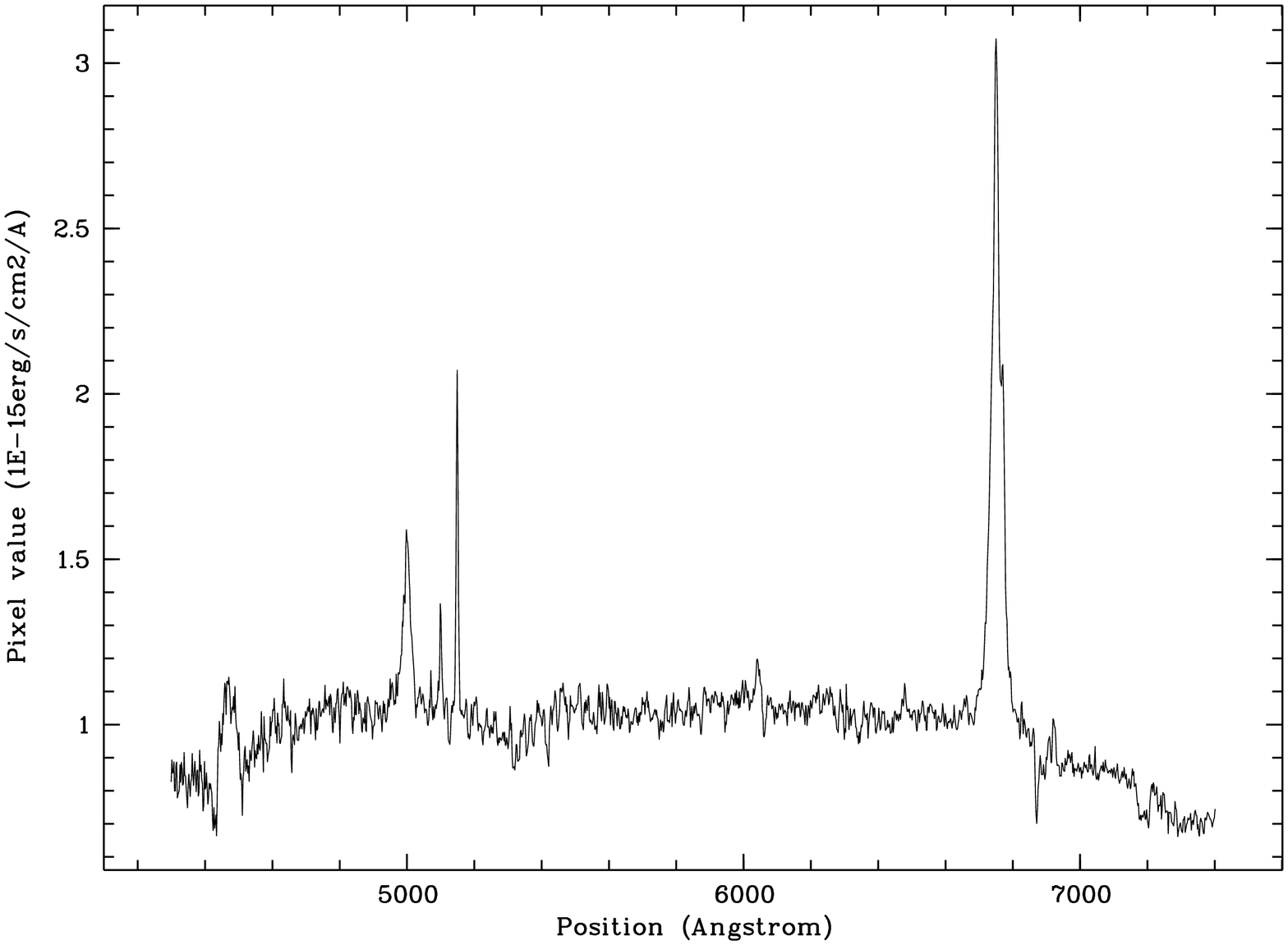,height=5.0cm,angle=-90,clip=}
      \vspace{0.5cm}
      \end{minipage}
      \hfill
      \begin{minipage}{6.0cm}
      \vspace{0.5cm}
      \centerline{CGCG 374-029}
      \psfig{figure=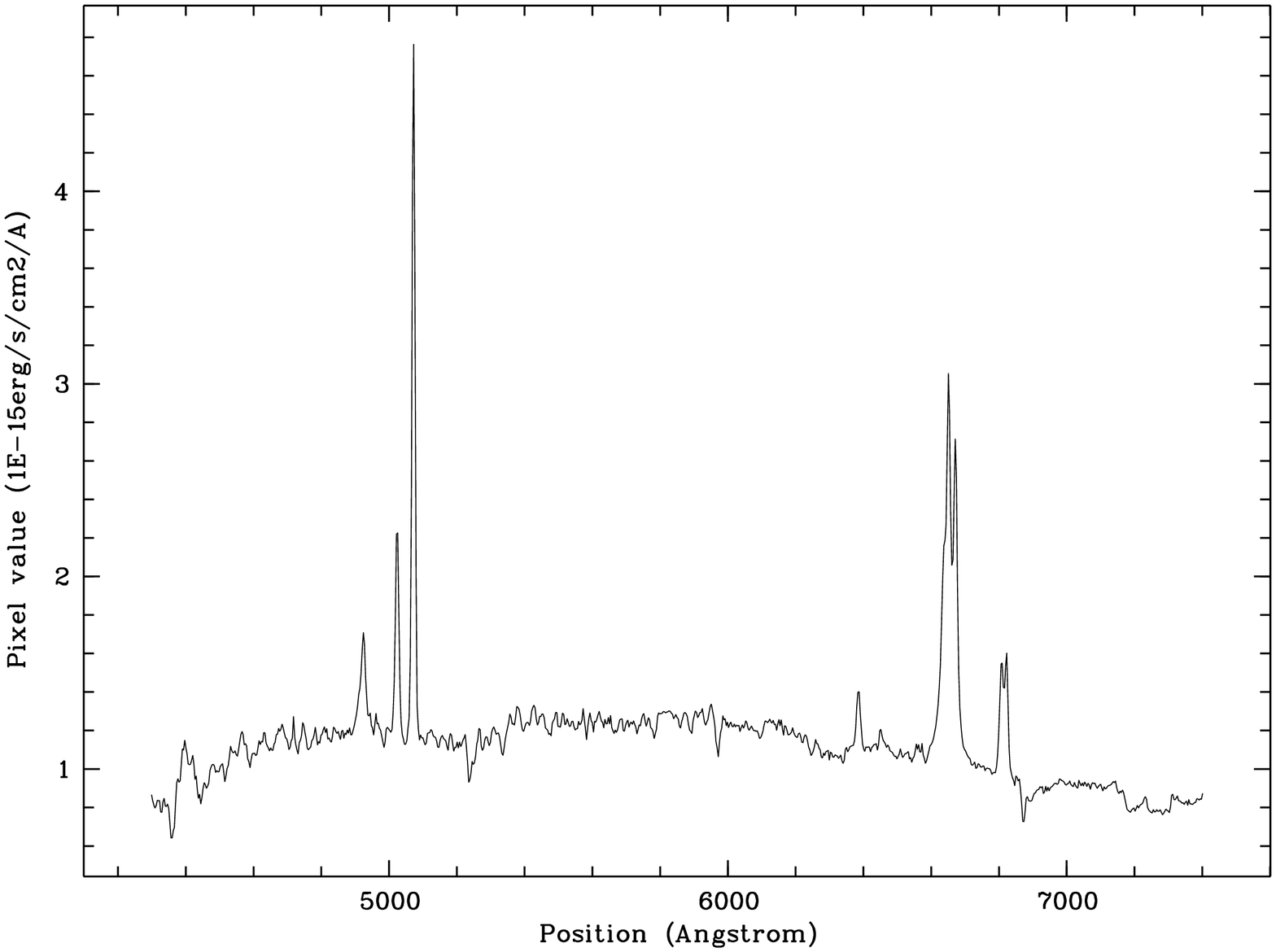,height=5.0cm,angle=-90,clip=}
      \vspace{0.5cm}
      \end{minipage}}
\hbox{\begin{minipage}{6.0cm}
      \centerline{CGCG 493-004}
      \psfig{figure=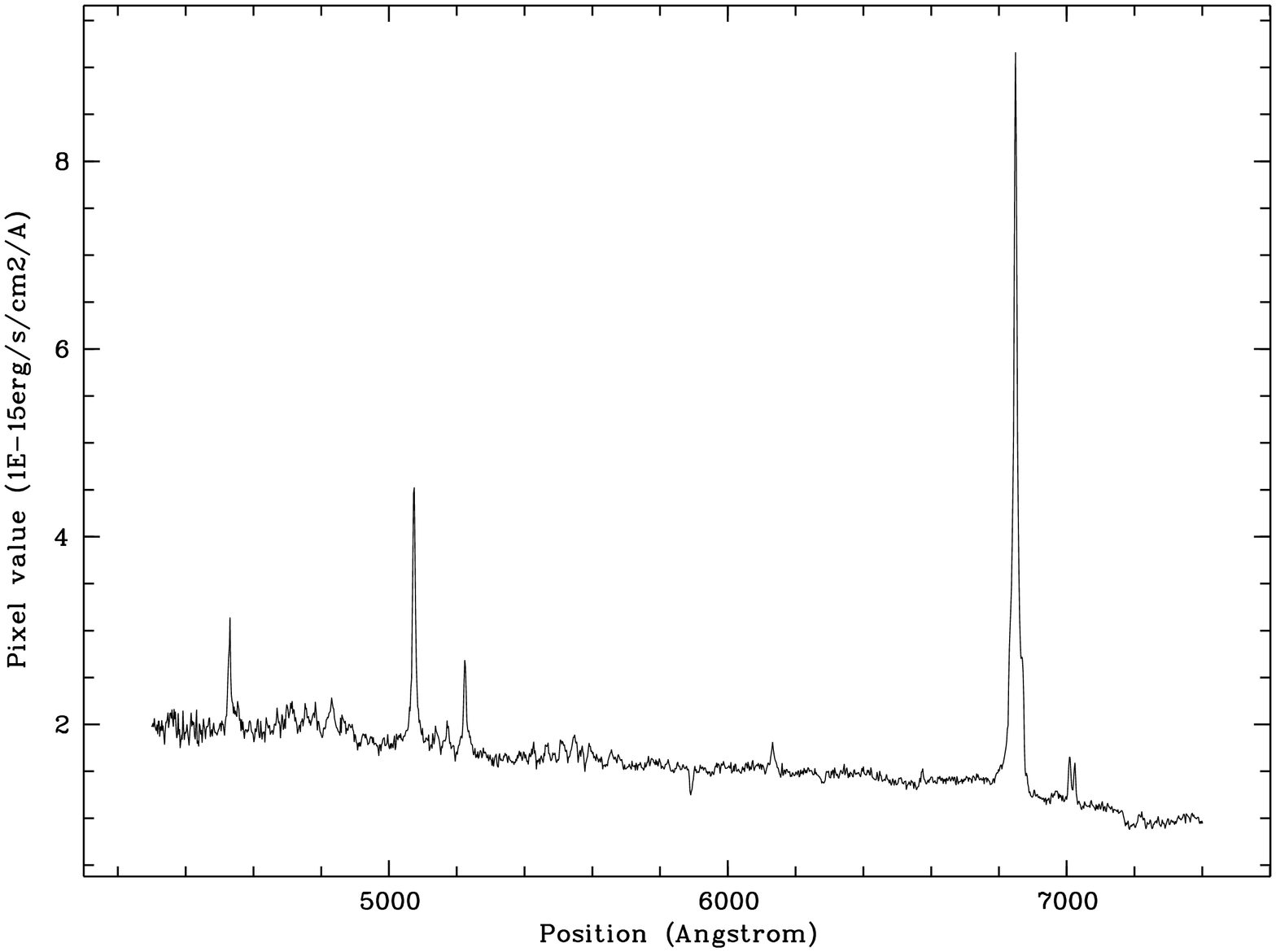,height=5.0cm,angle=-90,clip=}
      \vspace{0.5cm}
      \end{minipage}
      \hfill
      \begin{minipage}{6.0cm}
      \centerline{NGC 7158}
      \psfig{figure=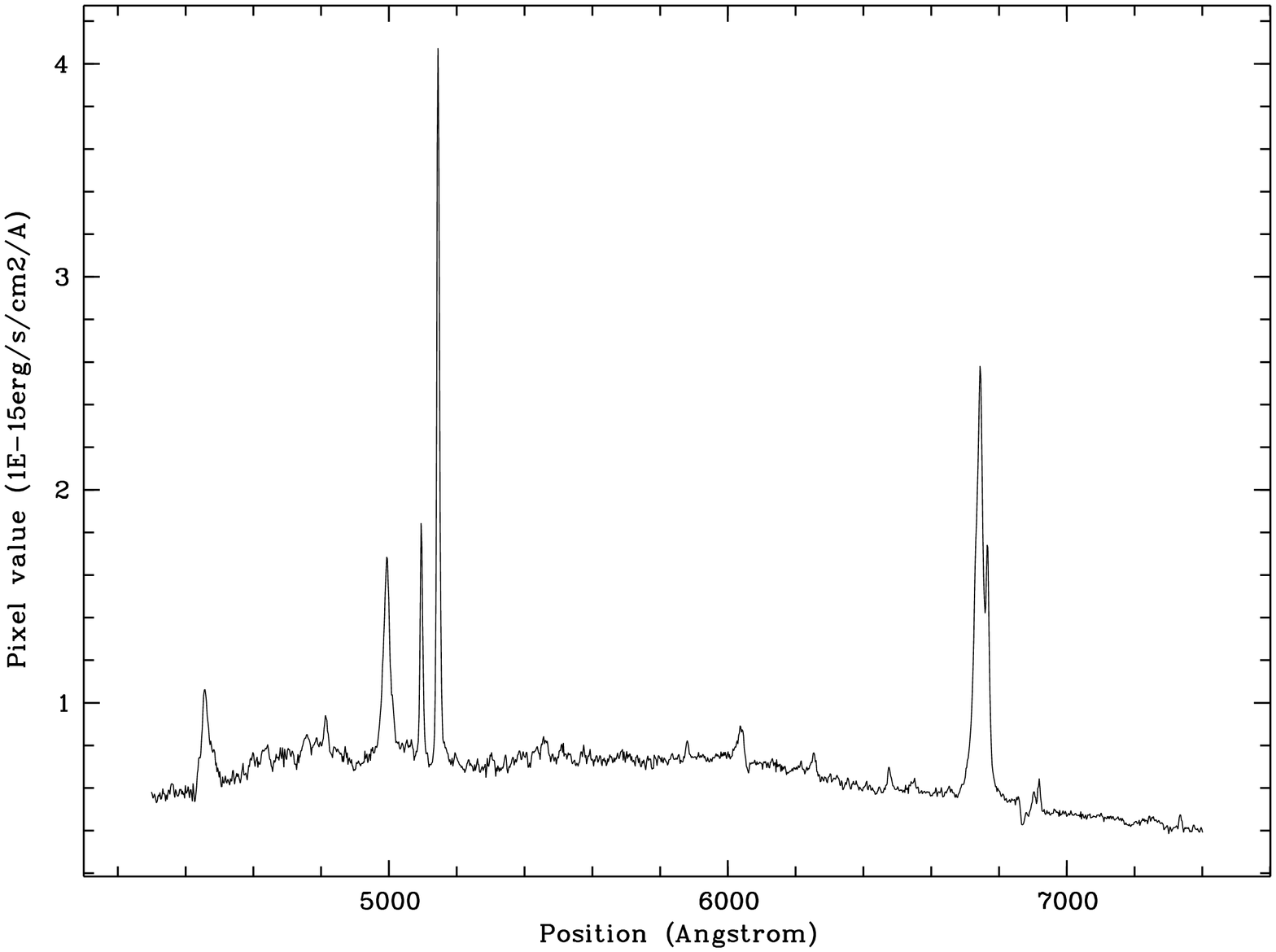,height=5.0cm,angle=-90,clip=}
      \vspace{0.5cm}
      \end{minipage}}
\hbox{\begin{minipage}{6.0cm}
      \centerline{II ZW 177}
      \psfig{figure=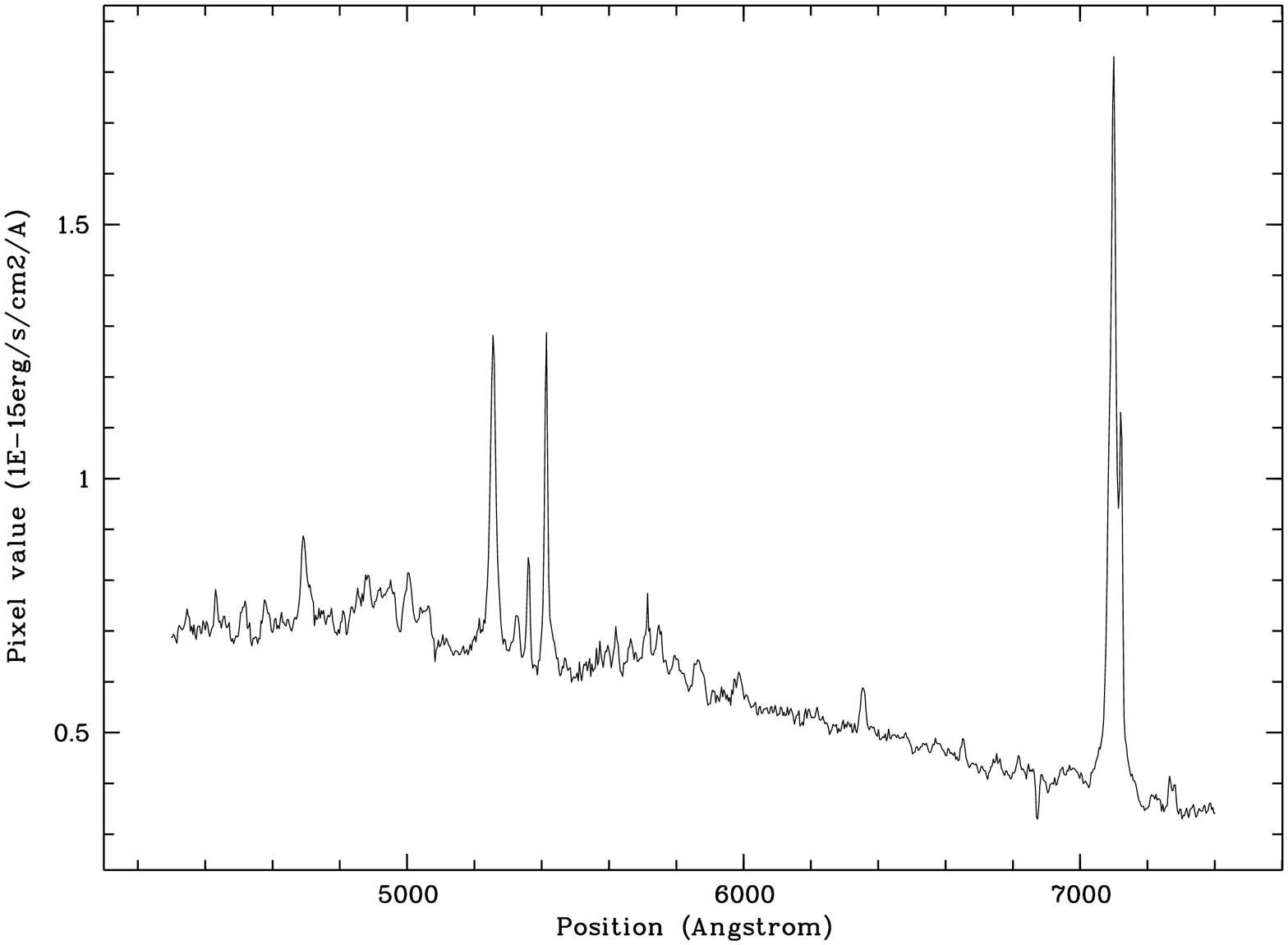,height=5.0cm,angle=-90,clip=}
      \vspace{0.5cm}
      \end{minipage}
      \hfill
      \begin{minipage}{6.0cm}
      \centerline{ESO 407-17}
      \psfig{figure=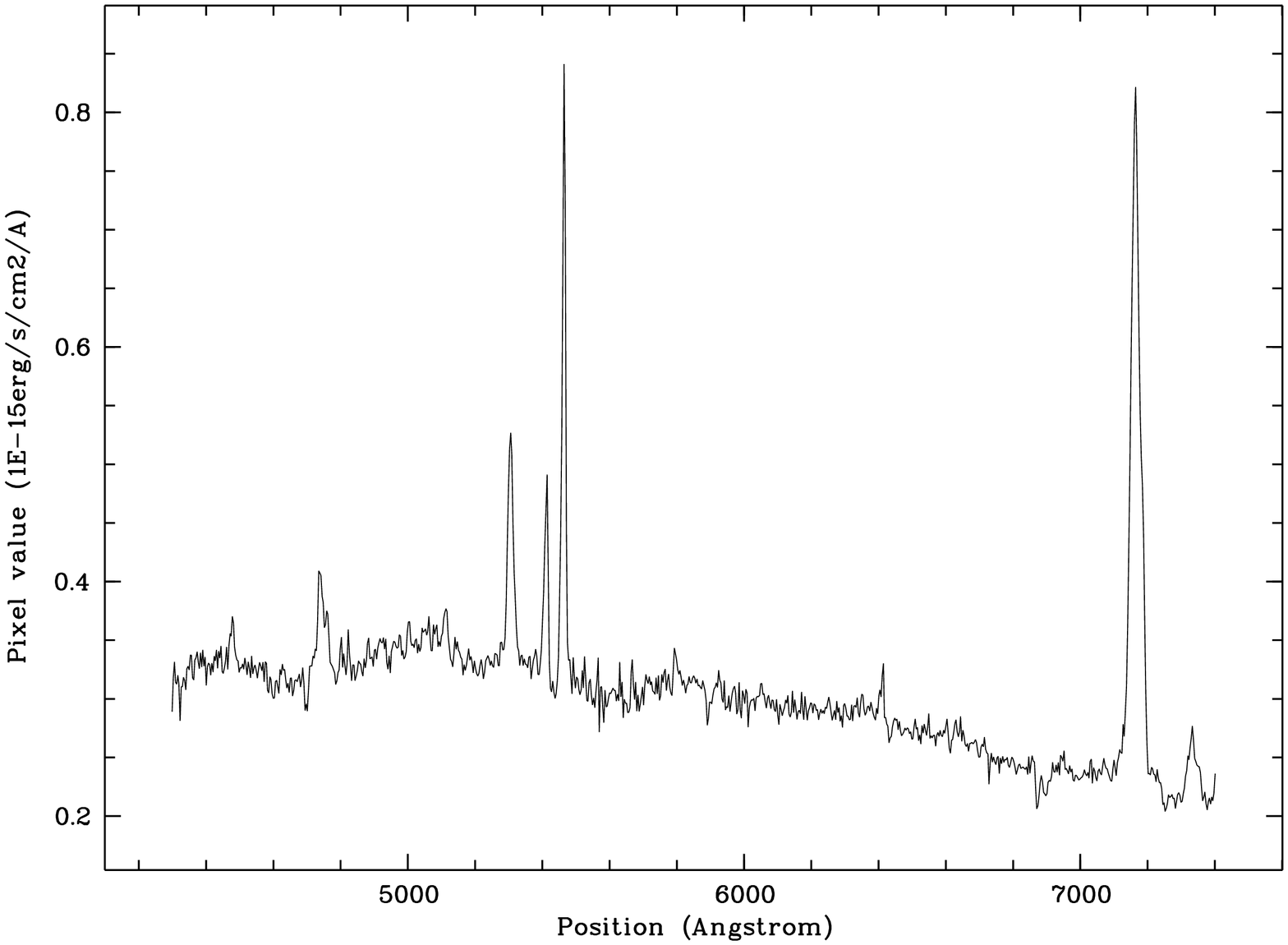,height=5.0cm,angle=-90,clip=}
      \vspace{0.5cm}
      \end{minipage}}
}
\caption{Optical spectra of RASS selected NLS1 nuclei taken at the 2.2 m telescopes
at Calar Alto and La Silla}
\end{figure}

\end{document}